\documentstyle[epsf,prd,aps]{revtex}

\newcommand{\be}{\begin{equation}}
\newcommand{\ee}{\end{equation}}

\title{Effective action for QED${}_4$ through $\zeta$-function regularization}

\bigskip

\author{C.~G.~Beneventano\thanks{Fellow FOMEC-UNLP (Argentina)}\and
E.~M.~Santangelo\thanks{Member of CONICET (Argentina)}}

\bigskip

\address{Departamento de F\'{\i}sica, Facultad de Ciencias Exactas,
Universidad Nacional de La Plata \\ C.C. 67 (1900) La Plata, Argentina}

\bigskip

\date{February 1, 2001}

\begin{document}
%\draft

\maketitle

\begin{abstract}

We obtain, through $\zeta$ function methods, the one-loop effective
action for massive Dirac fields in the presence of a uniform, but
otherwise general, electromagnetic background. After discussing
renormalization, we compare our $\zeta$ function result with
Schwinger's proper-time approach.

\end{abstract}

\pagebreak

\section{Introduction}
\label{sec-int}

In QED, the effective one-loop Lagrangian describes the effective
nonlinear interaction of the electromagnetic fields due to a
single fermion loop. In two dimensions, its general form has been
obtained both through proper time and $\zeta$ function
regularizations \cite{Schwinger2,gamboa}. In four dimensions, on
the other hand, only particular field configurations have been
studied.

The $3+1$ dimensional problem of constant electromagnetic fields was first
studied by Euler and Heisenberg \cite{Euler} and independently by
Weisskopf \cite{Weiss}. These authors obtained an integral expression for
the one-loop effective Lagrangian in the framework of the electron-hole
theory. Later on, Schwinger rederived this integral representation in a
field-theoretical scenario, by making use of proper time techniques
\cite{Schwinger}. In all these references, explicit results were derived
in some limits, the most famous being the weak-field one. This and other
particular field configurations were subsequently studied through the
proper-time regularization by a number of authors (see, for example,
\cite{Dittrich,Heyl,Dunne,Miel,Cho}).

More recently, the interest in the subject was renewed, and the
Euclidean effective action for constant electromagnetic background
configurations was studied through $\zeta$ function techniques
\cite{Dowker:1976tf,zeta}: In reference \cite{Blau} analytic
expressions were found for the case of purely magnetic fields in any
number of dimensions. In this same reference, the case of equal
electric and magnetic fields in four Euclidean dimensions was also
studied. A step towards more general field configurations was given
in \cite{Soldati}, where the authors obtained the effective
Lagrangian as a power series in $\frac{B}{E}$.

It is the aim of this paper to obtain, through $\zeta$ function
methods, an explicit non-perturbative expression for the full
one-loop effective action of Quantum Electrodynamics in four
dimensions in the case of constant, but otherwise arbitrary,
electromagnetic fields. To this end, we will work in Euclidean
space-time, and define the determinant of the relevant Dirac
operator $\not\!\!D$ through the derivative of the $\zeta$ function
of $\not\!\!D^{\dag}\not\!\!D$.

The organization of the paper is as follows:

After summarizing some well-known generalities in section
\ref{sec-setting}, we devote section \ref{sec-zeta} to analytically
extending the relevant $\zeta$ function to the region $\Re s>-2$.
(The main point here is the analytic extension of a Barnes
$\zeta$-function). Its value at $s=0$ is also given in this section.

In section \ref{seff}, a complete analytical expression for the
effective action in terms of special functions is given, and the
renormalization issue is discussed.

Section \ref{proptime} contains a comparison between $\zeta$ and
proper-time regularizations.

The Appendices \ref{lim-zeta} and \ref{lim-sef} contain the
derivation of some particular limits for the relevant zeta and for
the effective action, thus allowing for the comparison with previous
work on less general field configurations.

\section{Generalities}
\label{sec-setting}

We study the effective action for massive Dirac particles in the
presence of uniform, but otherwise arbitrary, electromagnetic
background fields. We work in four-dimensional Euclidean space.
Then, the effective action in the one-loop approximation is given
by
\begin{equation}
S_{eff}\left[A_{\mu}\right] = S_{cl}\left[A_{\mu}\right] - \log
Det\left(\not\!\!D\left[A_{\mu}\right]\right)\,,
\end{equation}
where $S_{cl}\left[A_{\mu}\right]$ is the classical Euclidean action
and $\not\!\!D\left[A_{\mu}\right] =
\gamma_{\mu}\left(\partial_{\mu} - ieA_{\mu}\right)+im$  is the
Euclidean Dirac operator, $m$ being the fermion mass.

Note that, even though $\not\!\!D$ is not self adjoint, it is
normal; so, the functional determinant appearing in the one-loop
correction to the action can be defined through $\zeta$ function
regularization \cite{Dowker:1976tf,zeta}, which leads to

\begin{equation}
S_{eff}\left[A_{\mu}\right] = S_{cl}\left[A_{\mu}\right]+
S^{(1)}\left[A_{\mu}\right]= S_{cl}\left[A_{\mu}\right] +\frac12
\left.\frac{\partial}{\partial s}
\zeta\left(s;\not\!\!D^{\dag}\not\!\!D\right)\right\rfloor_{s=0}\,.
\label{s}
\end{equation}

In order to evaluate the one-loop correction $S^{(1)}$ in the
previous expression, it is necessary to obtain the spectrum of the
operator $\not\!\!D^{\dag}\not\!\!D$, which is well known in the
case of uniform fields \cite{Bassetto}. In this particular
situation, one can always choose a reference frame such that
$F_{03}=-F_{30}=E$ and $F_{12}=-F_{21}=B$, while the remaining
components of the field tensor vanish. When doing so, the required
zeta function turns out to be
\[
\zeta\left(s;\not\!\!D^{\dag}\not\!\!D\right)={\mu}^4 \Omega
\frac{ab}{4{\pi}^2}\left[ 2\sum_{n_a=1}^{\infty} \left(2n_a
a+c\right)^{-s} + 2\sum_{n_b=1}^{\infty} \left(2n_b
b+c\right)^{-s}+\right.\]
\be
\left.4\sum_{n_a=1}^{\infty}\sum_{n_b=1}^{\infty} \left(2n_a
a+2n_b b+c\right)^{-s}+ c^{-s}\right]\,.\label{zeta0} \ee

Here, $\Omega$ is the volume of the four-dimensional Euclidean space,
$a=\frac{e|E|}{{\mu}^2}$, $b=\frac{e|B|}{{\mu}^2}$,
$c=\frac{m^2}{{\mu}^2}$, and $\mu$ is a parameter with mass
dimension, introduced to render the $\zeta$ function dimensionless.
Note that the series in equation (\ref{zeta0}) are all convergent
for $\Re s>2$, where they define an analytic function of $s$.

\section{Analytic extension of the $\zeta$ function}
\label{sec-zeta}

In this section, we will perform the analytic extension of the
relevant $\zeta$ function to a region containing $s=0$. In
particular, we will show it to be finite at $s=0$ and give its value
at this point.

The first two terms in equation (\ref{zeta0}) can be rewritten in
terms of Hurwitz' zeta functions, which are well known to be
meromorphic functions with a unique simple pole at $s=1$. On the
other hand, the third term is a zeta function of the Barnes' type
\cite{Barnes1,Barnes2} (see also \cite{Bordag2:1996,Klausmc} and
references therein). In order to analytically extend this term, we
write it in integral form. After doing so, we get

\bigskip

\[
\zeta\left(s;\not\!\!D^{\dag}\not\!\!D\right)={\mu}^4 \Omega
\frac{ab}{4{\pi}^2}\left\{\frac{2}{\left(2a\right)^s}\,
\zeta\left(s,\frac{c}{2a}+1\right)+
\frac{2}{\left(2b\right)^s}\,\zeta\left(s,\frac{c}{2b}+1\right)+\right.\]
\be \left.\frac{1}{\Gamma(s)} \int_{0}^{\infty} dt\,
t^{s-1}\frac{4e^{-2at}e^{-2bt}e^{-ct}}{\left(1-e^{-2at}\right)\left(1-e^{-2bt}\right)}\,+
c^{-s}\right\}\,=\,{\rm A}(s)+{\rm B}(s)+{\rm C}(s)+{\rm D}(s)\,,
\label{zeta} \ee where $\zeta(s,v)$ is Hurwitz' zeta function. This
expression (invariant under $a \leftrightarrow b$) is, in principle,
well defined for $\Re s>2$. Since the analytic structure of ${\rm
A}(s)$ and ${\rm B}(s)$ is well known, we will concentrate on the
Barnes term ${\rm C}(s)$, which will be extended to $\Re s>-2$.

To this end, we will use the expansion \cite{rusa} \be
\frac{1}{e^{at}-e^{-at}}= \frac{1}{2at}+at\sum_{k=1}^{\infty}(-1)^k
\frac{1}{(at)^2 +(k\pi)^2}\,,\label{sum} \ee thus obtaining
\[
{\rm C}(s)=2{\mu}^4 \Omega \frac{a
b}{4{\pi}^2}\frac{1}{\Gamma(s)}\left\{\frac{1}{2a} \int_{0}^{\infty}
dt\, t^{s-2} \frac{e^{-(a+b+c)t}} {e^{bt}-e^{-bt}}+\right.\]
\[ \left. a
\int_{0}^{\infty} dt\, t^{s} \frac{e^{-(a+b+c)t}}
{e^{bt}-e^{-bt}}\sum_{k=1}^{\infty}(-1)^k \frac{1}{(a t)^2
+(k\pi)^2}\right\} \, + \, a\leftrightarrow b\, =\] \be {\rm C}_{1}
(s)+{\rm C}_{2} (s)\,.\label{c} \ee

\bigskip

The first term, ${\rm C}_1 (s)$, can be easily seen to be \be {\rm
C}_1 (s)= 2{\mu}^4 \Omega
\frac{ab}{4{\pi}^2}\frac{1}{2a}\frac{1}{(s-1)\left(2b\right)^{s-1}}
\zeta\left(s-1,\frac{a+2b+c}{2b}\right) \, +\,a\leftrightarrow b
\,.\label{c1} \ee

As all the terms we have analytically extended up to this point,
${\rm C}_{2} (s)$ in equation (\ref{c} )involves an integral which
diverges at $s=0$. In order to isolate this singularity, we will
rewrite this term as
\[
{\rm C}_{2} (s)= 2{\mu}^4 \Omega \frac{a
b}{4{\pi}^2}\frac{1}{\Gamma(s)}\,a \int_{0}^{\infty} dt\, t^{s}
\frac{e^{-(a+b+c)t}} {\left(e^{bt}-e^{-bt}\right)}
\left\{\sum_{k=1}^{\infty}(-1)^k\left[\frac{1}{(a t)^2
+(k\pi)^2}-\frac{1}{(k\pi)^2}\right]\,+\right.
\]
\be \left.\sum_{k=1}^{\infty}(-1)^k \frac{1}{(k\pi)^2}\right\} \, +
\,a\leftrightarrow b\, ={\rm CF}_{2} (s)+{\rm CD}_2 (s)\,.\ee

The integral appearing in ${\rm CD}_{2} (s)$ is divergent at $s=0$
but, after performing the sum, it can be trivially extended to give
\be {\rm CD}_{2} (s)= -{\mu}^4 \Omega
\frac{ab}{4{\pi}^2}\frac{a}{6}\frac{s}{\left(2b\right)^{s+1}}
\zeta\left(s+1,1+\frac{a+c}{2b}\right)\, + \,a\leftrightarrow b
\,.\label{cd2} \ee

Now, ${\rm CF}_{2} (s)$, can be rewritten as \be {\rm CF}_{2} (s)=
-2{\mu}^4 \Omega \frac{a b}{4{\pi}^2}\frac{1}{\Gamma(s)}\,a^3
\sum_{k=1}^{\infty}\frac{(-1)^{k}}{(k\pi)^2} \int_{0}^{\infty} dt\,
t^{s+2} \frac{e^{-(a+2b+c)t}} {\left(1-e^{-2bt}\right)}
\frac{1}{(at)^2 +(k\pi)^2}+ \,a\leftrightarrow b\,. \label{fin}\ee

As is easily seen, this integral converges for $\Re s>-2$. We have
thus obtained an analytic extension for the $\zeta$ of the operator
as a meromorphic function with only simple poles. Such extension is
valid for $\Re s>-2$.

Now, the factor $\frac{1}{(a t)^2 +(k\pi)^2}$ can be written as an
integral. In fact
\[
\frac{1}{(a t)^2
+(k\pi)^2}=\frac{-1}{2ik\pi}\left[\frac{1}{at+ik\pi}-\frac{1}{at-ik\pi}\right]=
\frac{1}{k\pi}\int_0^{\infty}du\,e^{-atu}\sin(k\pi u)\,.\]

When replaced in equation (\ref{fin}), this gives
\[
{\rm CF}_{2}(s)= -2{\mu}^4 \Omega \frac{a
b}{4{\pi}^2}\frac{1}{\Gamma(s)}\,a^3
\sum_{k=1}^{\infty}\frac{(-1)^{k}}{(k\pi)^3} \int_{0}^{\infty} dt\,
t^{s+2} \frac{e^{-(a+2b+c)t}}
{\left(1-e^{-2bt}\right)}\int_0^{\infty}du\,e^{-atu}\sin(k\pi u)+
\,a\leftrightarrow b\] or, after interchanging the integrals
\[
{\rm CF}_{2}^2 (s)= -2{\mu}^4 \Omega \frac{a
b}{4{\pi}^2}\frac{a^3}{\Gamma(s)}
\sum_{k=1}^{\infty}\frac{(-1)^{k}}{(k\pi)^3} \int_{0}^{\infty}
du\,\sin(k\pi u)\frac{\Gamma(s+3)}{(2b)^{s+3}}
\zeta\left(s+3,\frac{a+2b+c+au}{2b}\right) + \,a\leftrightarrow
b\,.\]

When the $\zeta$ function is written in terms of its series
development (which is valid for $\Re s>-2$)  one has (after
interchanging this series and the integral)
\[
{\rm CF}_{2}(s)= -2{\mu}^4 \Omega \frac{a
b}{4{\pi}^2}\frac{a^3}{\Gamma(s)}\frac{\Gamma(s+3)}{(2b)^{s+3}}
\sum_{k=1}^{\infty}\frac{(-1)^{k}}{(k\pi)^3}
\sum_{l=1}^{\infty}\int_{0}^{\infty} du\,\sin(k\pi u)
\left(l+\frac{a+c+au}{2b}\right)^{-(s+3)} + \,a\leftrightarrow b \,.
\]

Finally, after performing the remaining integral and making use of
the functional relations between incomplete gamma functions
\cite{Abram}, one gets
\[
{\rm CF}_{2}(s)= i{\mu}^4 \Omega \frac{a
b}{4{\pi}^2}\frac{\Gamma(s+3)}{\Gamma(s)}\,a^{-s}\frac{1}{s+2}
\sum_{k=1}^{\infty}\frac{(-1)^{k}}{(k\pi)^{1-s}}
\sum_{l=1}^{\infty}\left[i^{s+2}e^{i\frac{k\pi}{a}(2bl+a+c)}
\Gamma\left(-s-1,i\frac{k\pi}{a}(2bl+a+c)\right)-\right.\] \be \left.
(-i)^{s+2}e^{-i\frac{k\pi}{a}(2bl+a+c)}
\Gamma\left(-s-1,-i\frac{k\pi}{a}(2bl+a+c)\right)\right]\, +
\,a\leftrightarrow b \,.\label{cf2}\ee

The replacement of equations (\ref{c1}), (\ref{cd2}) and (\ref{cf2})
into equation (\ref{zeta}) completes the analytic extension of the
relevant $\zeta$ function. Its  value at $s=0$ can be easily
computed, which gives:

\be \zeta\left(0;\not\!\!D^{\dag}\not\!\!D\right)= \frac{{\mu}^4
\Omega}{4{\pi}^2}\left\{\frac12 c^2 + \frac{a^2 +b^2}{3}\right\}\,.
\label{zs0}\ee

The agreement with the known results for null and equal fields is
shown in Appendix \ref{lim-zeta}.

\section{The effective action and its renormalization}
\label{seff}

This section contains the main result in this paper, i.e., the
one-loop correction to the Euclidean effective action. According
to equation (\ref{s}), to obtain such result, one must perform the
derivatives at $s=0$ of the various terms in equation
(\ref{zeta}).

We start from ${\rm A}(s)$, which contributes with
\be
\frac12\left. \frac{\partial}{\partial s}{\rm
A}(s)\right\rfloor_{s=0}= {\mu}^4 \Omega
\frac{ab}{4{\pi}^2}\left\{\log(2a) \left(\frac12
+\frac{c}{2a}\right)+\log \Gamma(\frac{c}{2a}+1)-\frac12
\log(2\pi)\right\}\,.\label{dera} \ee

In a completely analogous way, one has
\be
\frac12\left. \frac{\partial}{\partial s}{\rm
B}(s)\right\rfloor_{s=0}= {\mu}^4 \Omega
\frac{ab}{4{\pi}^2}\left\{\log(2b) \left(\frac12
+\frac{c}{2b}\right)+\log \Gamma(\frac{c}{2b}+1)-\frac12
\log(2\pi)\right\}\,.\label{derb} \ee

It is also through a direct calculation that one gets \be \frac12
\left.\frac{\partial}{\partial s}{\rm C}_1(s)\right\rfloor_{s=0}=
{\mu}^4 \Omega \frac{ab}{4{\pi}^2}\frac{1}{2a}\left\{2b\left(-1+
\log(2b)\right) \zeta(-1,1+\frac{a+c}{2b}) -2b
\left.\frac{\partial}{\partial s} \right\rfloor_{s=0}
\zeta(s-1,1+\frac{a+c}{2b})\right\}+\,a\leftrightarrow b \,.
\label{derc1}\ee

\be
\frac12 \left.\frac{\partial}{\partial s}{\rm
CD}_2(s)\right\rfloor_{s=0}= {\mu}^4 \Omega
\frac{ab}{4{\pi}^2}\frac{a}{24b}\left\{\log(2b) + \Psi
(1+\frac{a+c}{2b})\right\} +\,a\leftrightarrow b \,.\label{dercd2}\ee

As regards ${\rm CF}_2(s)$, due to the presence of $\Gamma(s)$ in
the denominator, the required derivative reduces to the product
$\Gamma(s)\,{\rm CF}_2(s)$ at $s=0$, i.e.,
\[
\frac12 \left.\frac{\partial}{\partial s}{\rm
CF}_2(s)\right\rfloor_{s=0}= -\frac{i}{2}{\mu}^4 \Omega \frac{a
b}{4{\pi}^2} \sum_{k=1}^{\infty}\frac{(-1)^{k}}{k\pi}
\sum_{l=1}^{\infty}\left[e^{i\frac{k\pi}{a}(2bl+a+c)}
\Gamma\left(-1,\frac{i k\pi}{a}(2bl+a+c)\right)-\right.
\]
\be \left.e^{-i\frac{k\pi}{a}(2bl+a+c)} \Gamma\left(-1,-\frac{i
k\pi}{a}(2bl+a+c)\right)\right] + \,a\leftrightarrow b \,.
\label{dercf2}\ee

Summarizing, the Euclidean effective action is given by the sum of
the partial contributions in equations (\ref{dera}) to
(\ref{dercf2}), plus \be \frac12 \left.\frac{\partial}{\partial
s}{\rm D}(s)\right\rfloor_{s=0}= -{\mu}^4 \Omega \frac{a
b}{8{\pi}^2}\log(c)\,.\label{derd} \ee

Notice that even though the result is finite, it depends on the
arbitrary parameter $\mu$. However, this effective action still
admits a finite renormalization. We will perform it by adopting the
criterium (used, for instance, in reference \cite{tubo}), that a very
massive field does not fluctuate. Thus, we will subtract the one loop
correction to the effective action in the limit
$m\rightarrow\infty$, $\mu \rightarrow\infty$, with constant $c$.
From equation (\ref{lims}) in Appendix \ref{lim-sef}, the effective
action in this limit can be seen to be \be {\mu}^4 \Omega
\frac{1}{4{\pi}^2}\left\{\left[\frac38 -\frac14 \log(c)\right]c^2 -
\frac16 (b^2 +a^2)\log(c)\right\}\,. \ee

After doing this subtraction, all dependence on the parameter $\mu$
disappears, and the Euclidean effective action is given by
\[
S_{eff}^{Ren}\left[A_{\mu}\right] =
\frac{\Omega{\mu}^4}{2e^2}(a^2+b^2)\,+\]
\[{\mu}^4 \Omega
\frac{ab}{4{\pi}^2}\left\{\frac18
\log\left(\frac{4ab}{c^2}\right)-\frac{1}{24}\frac{(a^2
+b^2)}{ab}\log\left(\frac{4ab}{c^2}\right)+\frac{c}{4a}\log\left(\frac{a}{b}\right)-
\frac{c^2}{16 ab}\log\left(\frac{4ab}{c^2}\right)+\right.\]
\[
\log\left(\frac{\Gamma(\frac{c}{2a}+1)}{\sqrt{2 \pi}}\right)-
\frac{b}{a}\zeta\left(-1,1+\frac{a+c}{2b}\right)-\frac{b}{a}\left.\frac{\partial}{\partial
s}\right\rfloor_{s=0}\zeta\left(s-1,1+\frac{a+c}{2b}\right)-
\]
\[\frac{i}{2}\sum_{k=1}^{\infty}\frac{(-1)^{k}}{k\pi}
\sum_{l=1}^{\infty}\left[e^{i\frac{k\pi}{a}(2bl+a+c)}
\Gamma\left(-1,\frac{i
k\pi}{a}(2bl+a+c)\right)-e^{-i\frac{k\pi}{a}(2bl+a+c)}
\Gamma\left(-1,\frac{-i k\pi}{a}(2bl+a+c)\right)\right]+\]
\be\left.\frac{a}{24 b} \Psi\left(1+\frac{a+c}{2b}\right)-
\frac{3}{16} \frac{c^2}{ab}\right\}\,+ \,a\leftrightarrow b\,.
\label{sren}\ee

The renormalization performed amounts to subtracting the zero field
effective action (thus redefining the cosmological constant), and
renormalizing the classical action. As a result, one gets the
following running charge relationship \be
\frac{1}{e^2}=\frac{1}{e_{0}^2}+\frac{1}{12\pi^2}\log
\frac{\mu^2}{m^2}\,.\ee

Equivalently, for the fine structure constant one has \be
\alpha=\frac{\alpha_0}{1+\frac{\alpha_0 }{3\pi} \log
\frac{\mu^2}{m^2}}\,.\ee

Note that this expression reduces, in the perturbative limit, to the
well known result (see, for example \cite{itzykson}) \be
\alpha=\alpha_0 (1-\frac{\alpha_0 }{3\pi} \log
\frac{\mu^2}{m^2})\,.\ee

\section{Comparison with the proper time result}
\label{proptime}

In Appendix \ref{lim-sef} we show that, in the weak field limit, our
result for the $\zeta$ regularized effective action coincides, once
renormalized, with the Euclidean version of the well known
Schwinger's proper time one.

In this section, we will show that this is also the case for
arbitrary field strenghts. In fact, Schwinger's integral expresion
for the one loop correction to the effective action is given, after
subtracting the divergent terms, by \be S_{PT}^{(1)}={\mu}^4 \Omega
\left.\left\{\frac{ab}{8{\pi}^2}\int_{0}^{\infty} dt\,
t^{s-1}e^{-ct}\coth(bt)\coth(at)-
\frac{1}{8{\pi}^2}\int_{0}^{\infty} dt\, t^{s-3}e^{-ct}-
\frac{a^2+b^2}{24{\pi}^2}\int_{0}^{\infty} dt\,
t^{s-1}e^{-ct}\right\}\right\rfloor_{s=0}\,.\ee

Now, performing the integrals in the last two terms and comparing
with equation (\ref{zeta}) (with the Hurwitz's zetas written in
integral form), the previous expression can be rewritten as \be
S_{PT}^{(1)}=\frac12 \left.\left\{\Gamma(s)
\zeta\left(s;\not\!\!D^{\dag}\not\!\!D\right)-\frac{{\mu}^4
\Omega}{4{\pi}^2}\left(c^{2-s}\Gamma(s-2)+
\frac{a^2+b^2}{3}c^{-s}\Gamma(s)\right)\right\}\right\rfloor_{s=0}\,.\ee

After developing around $s=0$, it is easy to see that \be S_{PT}^{(1)}=
S_{\zeta}^{(1)} -\frac{{\mu}^4
\Omega}{4{\pi}^2}\left[\frac{3}{8}c^2-\left(\frac{c^2}{4}+\frac{a^2+b^2}{6}
\right)\log c\right]\,,\ee where $S_{\zeta}^{(1)}$ is the
$\zeta$-regularized one loop correction to the effective action, as
defined in equation (\ref{s}), and the remaining terms are precisely the
ones we have subtracted through renormalization. So, the exact agreement
between both renormalized effective actions is apparent.

\acknowledgements

We thank Horacio Falomir, Klaus Kirsten and Roberto Soldati for
carefully reading the manuscript, and for many useful suggestions.
This work was partially supported by UNLP, under Grant No 11/X230,
ANPCyT, under Grant PICT00039, and CONICET, under Grant PIP0459.

\appendix

\section{The limits of null and equal fields}
\label{lim-zeta}

In this section, we will show the agreement of our general $\zeta$
function with the results obtained by other authors for some
particular cases, i.e., the case of a null electric or magnetic
field \cite{Blau,Soldati} and that of equal electric and magnetic
fields \cite{Blau}.

We will start with the $B\rightarrow 0$ limit. It is easy to see
that $\lim_{b\rightarrow 0}{\rm A}(s)=0$. As regards
$\lim_{b\rightarrow 0}{\rm B}(s)$, it can be studied by making use
of the asymptotic expansion for Hurwitz' $\zeta$ function (see, for
example, \cite{Bateman}) \be
\zeta(s,v)=\frac{1}{\Gamma(s)}\left\{v^{1-s}\Gamma(s-1)+ \frac12
v^{-s}\Gamma(s)+ \sum_{n=1}^{N}
B_{2n}\frac{\Gamma(s+2n-1)}{(2n)!}v^{1-s-2n}\right\}+
O(v^{-2N-s-1})\,, \label{desas}\ee which gives \be
\lim_{b\rightarrow 0}{\rm B}(s)=\lim_{b\rightarrow 0} \left\{{\mu}^4
\Omega
\frac{ab}{4{\pi}^2}\frac{2}{(2b)^s}\frac{\Gamma(s-1)}{\Gamma(s)}
\left(\frac{c}{2b}+1\right)^{1-s}\right\}= \frac{{\mu}^4
\Omega}{4{\pi}^2}\frac{a}{s-1}c^{1-s}\,.\ee

The only contribution to ${\rm C}(s)$ in this limit comes from ${\rm
C}_1(s)$, which gives \be \lim_{b\rightarrow 0}{\rm C}(s)=
\frac{{\mu}^4
\Omega}{4{\pi}^2}\frac{(2a)^{2-s}}{s-1}\left\{\zeta(s-1,\frac{c}{2a})-
\left(\frac{c}{2a}\right)^{1-s}\right\}\,.\ee

Finally, ${\rm D}(s)$ vanishes for $b=0$. Then, replacing all these
partial results into equation (\ref{zeta}), one obtains \be
\zeta(s,\not\!\!D^{\dag}\not\!\!D)\rfloor_{b=0}= \frac{{\mu}^4
\Omega}{4{\pi}^2}\frac{(2)^{1-s}}{s-1}a^{2-s}\left\{2\zeta(s-1,\frac{c}{2a})-
\left(\frac{c}{2a}\right)^{1-s}\right\}\,\label{z}\ee which is in
complete agreement with previous results \cite{Blau,Soldati}.

Of course, the $E\rightarrow 0$ limit, gives an analogous
expression, which can be obtained by changing $a \rightarrow b$ in
equation (\ref{z}).
\bigskip

We will now study the equal fields limit. In this situation, taking
$a=b$ in the different terms appearing in the $\zeta$ function
(\ref{zeta}), we have
\[
\left.\zeta\left(s;\not\!\!D^{\dag}\not\!\!D\right)\right\rfloor _
{a=b}={\mu}^4 \Omega
\frac{a^2}{4{\pi}^2}\left\{\frac{4}{\left(2a\right)^s}\,
\zeta\left(s,\frac{c}{2a}+1\right)\,+ c^{-s}\, + \frac{2^{2-s} a^{-s}
}{s-1}\,\zeta\left(s-1,\frac32 +\frac{c}{2a}\right)\,-\right.\]
\[\frac16 (2a)^{-s}\,\zeta\left(s+1,\frac32 +\frac{c}{2a}\right)\,-
i \,2\,a^{-s}(s+1)s
\sum_{k=1}^{\infty}\frac{(-1)^{k+1}}{(k\pi)^{1-s}}
\sum_{l=1}^{\infty}\left[i^{s+2}e^{ik\pi (2l+1+\frac{c}{a})}
\Gamma\left(-s-1,ik\pi (2l+1+\frac{c}{a})\right)-\right.\] \be
\left.\left. (-i)^{s+2}e^{-ik\pi(2l+1+\frac{c}{a})}
\Gamma\left(-s-1,-ik\pi(2l+1+\frac{c}{a})\right)\right]\right\}\,.\ee

In order to compare this expression with the result in \cite{Blau},
we use the functional relations between incomplete gamma functions,
thus getting

\[
\left.\zeta\left(s;\not\!\!D^{\dag}\not\!\!D\right)\right\rfloor _
{a=b}={\mu}^4 \Omega
\frac{a^2}{4{\pi}^2}\left\{\frac{4}{\left(2a\right)^s}\,
\zeta\left(s,\frac{c}{2a}+1\right)\,+ c^{-s}\, + \frac{2^{2-s} a^{-s}
}{s-1}\,\zeta\left(s-1,\frac32 +\frac{c}{2a}\right)\,-\right.\]
\[i\,2\, a^{-s} s\sum_{k=1}^{\infty}\frac{(-1)^{k}}{(k\pi)^{1-s}}
\sum_{l=1}^{\infty}\left[i^{s+2}e^{ik\pi (2l+1+\frac{c}{a})}
\Gamma\left(-s,ik\pi (2l+1+\frac{c}{a})\right)-\right.\]
\be\left.\left. (-i)^{s+2}e^{-ik\pi(2l+1+\frac{c}{a})}
\Gamma\left(-s,-ik\pi(2l+1+\frac{c}{a})\right)\right]\right\}\,.\label{p1}\ee

We now use the integral representation for the incomplete gamma
function \[\Gamma(\alpha ,x)=\int_{x}^{\infty}dt\, e^{-t}
t^{\alpha-1}\,.\] When doing so, and after interchanging the
integral and the sum over $l$, the last term in equation(\ref{p1})
can be written as
\[(2a)^{-s}
s\sum_{k=1}^{\infty}\frac{(-1)^{k}}{(k\pi)^{2}}\int_{0}^{\infty}du\,
e^{-u}\left[\zeta\left(s+1, \frac32 +\frac{c}{2a}-\frac{i
u}{2k\pi}\right) + \zeta\left(s+1, \frac32 +\frac{c}{2a}+\frac{i
u}{2k\pi}\right)\right]=\]
\[2(2a)^{-s}\frac{1}{\Gamma(s)}
\sum_{k=1}^{\infty}(-1)^{k}\int_{0}^{\infty}dt\, t^s
\frac{e^{-(\frac32
+\frac{c}{2a})t}}{1-e^{-t}}\frac{1}{(k\pi)^{2}+(\frac t2)^2}
\] where we have used the integral form for the
Hurwitz's zeta functions, interchanged the integrals and performed
the interior one.

Interchanging now the integral with the sum, and using equation
(\ref{sum}), we obtain
\[2^{2-s} a^{-s} \frac{1}{\Gamma(s)}\int_{0}^{\infty}dt\, t^{s-1}
\frac{e^{-(\frac32 +\frac{c}{2a})t}}{1-e^{-t}}\left[\frac{e^{-\frac
t2}}{1-e^{-t}}\,-\frac1t\right]=
\]
\[2^{2-s} a^{-s}\left[\zeta\left(s-1,\frac{c}{2a}+1\right)-
(\frac{c}{2a}+1)\zeta\left(s,\frac{c}{2a}+1\right)-
\frac{1}{s-1}\zeta\left(s-1,\frac32 +\frac{c}{2a}\right)\right]\,.\]
When replaced in (\ref{p1}), the final result is

\[
\left.\zeta\left(s;\not\!\!D^{\dag}\not\!\!D\right)\right\rfloor _
{a=b}={\mu}^4 \Omega \frac{a^2}{4{\pi}^2}\left\{c^{-s}\,+ 2^{2-s}
a^{-s}\left[\zeta\left(s-1,\frac{c}{2a}+1\right) -
\frac{c}{2a}\zeta\left(s,\frac{c}{2a}+1\right)\right]\right\}=\]
\be{\mu}^4 \Omega \frac{a^2}{4{\pi}^2}\left\{c^{-s}\,+ 4
(2a)^{-s}\left(\zeta\left(s-1,\frac{c}{2a}\right)-
\frac{c}{2a}\zeta\left(s,\frac{c}{2a}\right)\right)\right\}\,.\ee
This expression coincides whith the result obtained in \cite{Blau}
(see equations (5.2.6) and (5.2.4) in that reference).

\bigskip

\section{The weak-field limit}
\label{lim-sef}

An unavoidable test our effective action must resist is its
coincidence whith the well known result for weak fields
\cite{Euler,Schwinger}. In order to check this is the case, we will
develop the different contributions to the effective action
(equations (\ref{dera}) to (\ref{derd})) in powers of the fields over
the squared mass. In the cases of equations (\ref{dera}) to
(\ref{dercd2}), such development can be obtained by making use of
the well known asymptotic expansions \cite{Bateman} for
$\log\Gamma(x)$, $\psi(x)$, and $\zeta(s,x)$ (see also our equation
(\ref{desas})). When doing so, and retaining terms up to the order
of squared fields over mass to the fourth, one gets, after a
straightforward though tedious calculation, \be \frac12\left.
\frac{\partial}{\partial s}{\rm A}(s)\right\rfloor_{s=0}\simeq
{\mu}^4 \Omega \frac{ab}{4{\pi}^2} \left\{\frac16 a c^{-1}+\frac12
\log(c)+\frac{1}{2a}\left(\log(c)-1\right) c\right\}\,.\label{lima}
\ee \be \frac12\left. \frac{\partial}{\partial s}{\rm
B}(s)\right\rfloor_{s=0}\simeq {\mu}^4 \Omega \frac{ab}{4{\pi}^2}
\left\{\frac16 b c^{-1}+\frac12
\log(c)+\frac{1}{2b}\left(\log(c)-1\right) c\right\}\,.\label{limb}
\ee
\[
\frac12 \left.\frac{\partial}{\partial s}{\rm
C}_1(s)\right\rfloor_{s=0}\simeq {\mu}^4 \Omega
\frac{ab}{4{\pi}^2}\frac{1}{ab}\left\{\left(\frac14-\frac14 \log c
+\frac18\right)c^2 + \left(\frac12 (a+b)-\frac12 (a+b)\log
c\right)c-\frac{5}{24}(a^2 +b^2)-\right.\] \be\left.\frac12 ab \log c
- \frac{1}{24}\left(5 b a^2 +5 a b^2 + a^3 + b^3\right)
c^{-1}+\left(\frac{1}{24} b^3 a+ \frac{1}{24} a^3 b+ \frac{1}{12}
b^2 a^2 + \frac{7}{1440} a^4 + \frac{7}{1440} b^4\right)
c^{-2}\right\}\,.\label{limc1}\ee
\[
\frac12 \left.\frac{\partial}{\partial s}{\rm
CD}_2(s)\right\rfloor_{s=0}\simeq {\mu}^4 \Omega
\frac{ab}{4{\pi}^2}\frac{1}{24}\left\{\left(\frac{a}{b}+\frac{b}{a}\right)\log
c + \left( a+b+\frac{a^2}{b} +\frac{b^2}{a}\right) c^{-1} -\right.\]
\be \left.\frac12 \left( 2 a^2+ 2 b^2 + \frac{a^3}{b}+ \frac{b^3}{a}
+\frac43 ba\right)c^{-2}\right\} \,.\label{limcd2}\ee

As regards $\frac12 \left.\frac{\partial}{\partial s}{\rm
CF}_2(s)\right\rfloor_{s=0}$, one has to use the asymptotic
expansions for the incomplete $\Gamma$ function and for the Hurwitz'
zeta functions (equation (\ref{desas})). After doing so, one obtains
\be \frac12 \left.\frac{\partial}{\partial s}{\rm
CF}_2(s)\right\rfloor_{s=0}\simeq {\mu}^4 \Omega \frac{ab}{4{\pi}^2}
\frac{7}{1440}\left(\frac{a^3}{b}+\frac{b^3}{a}\right)c^{-2}\,.
\label{limcf2}\ee

By summing up the contributions in equations (\ref{lima}) to
(\ref{limcf2}), plus the one coming from $\frac12
\left.\frac{\partial}{\partial s}{\rm D}(s)\right\rfloor_{s=0}$, the
one-loop correction to the effective action is seen to reduce, in
this weak-field limit, to \be S^{(1)}={\mu}^4 \Omega
\frac{1}{4{\pi}^2}\left\{\left[\frac38 -\frac14 \log(c)\right]c^2 -
\frac16 (b^2 +a^2)\log(c)+ \left[\frac{7}{90}(ab)^2 -\frac{1}{90}(a^2
+b^2)^2\right]c^{-2}\right\}\,.\label{lims}\ee

Now, renormalizing according to the criterium discussed in Section
\ref{seff}, one is left with \be S_{eff}=\frac{\Omega}{2}(B^2+E^2)+
\frac{\Omega e^4}{8{\pi}^2 m^4 }\left[\frac{7}{45}(EB)^2
-\frac{1}{45}(E^2+B^2)^2\right] \,,\label{wf}\ee

where the definitions of $a$, $b$ and $c$ given in the paragraph
following equation (\ref{zeta0}) were used.

The expression in (\ref{wf}) is precisely the Euclidean version of
the Euler-Heisenberg effective action for weak fields
\cite{Euler,Schwinger}.

\end{document}